\documentstyle[aps,eqsecnum,epsfig]{revtex}
\begin{document}

\draft

\author{B.~All\'es$^a$, S. Caracciolo$^b$, A. Pelissetto$^c$, M. Pepe$^a$}

\address{$^a$Dipartimento~di Fisica, Universit\`a di Milano-Bicocca
      and INFN Sezione di Milano, Milano, Italy}

\address{$^b$Scuola Normale Superiore and INFN Sezione di Pisa, Pisa,
Italy}

\address{$^c$Dipartimento di Fisica and INFN Sezione di Roma I,
          Universit\`a di Roma {\it ``La Sapienza''}, Roma, Italy}

\title{Four-loop contributions to long-distance quantities in the 
two-dimensional nonlinear $\sigma$--model on a square lattice: 
revised numerical estimates\footnote{Preprint Bicocca--FT--99--17.}}
\maketitle

\vskip 1cm

\centerline{\it Abstract}
\begin{abstract}
We give the correct analytic expression of a finite integral appearing 
in the four-loop computation of the renormalization-group
functions for the two-dimensional nonlinear $\sigma$--model on the square 
lattice with standard action, explaining the origin of a numerical discrepancy.
We revise the numerical expressions of Caracciolo and Pelissetto 
for the perturbative corrections
of the susceptibility and of the correlation length. 
For the values used in Monte Carlo simulations, $N=3$, 4, 8, 
the second perturbative correction coefficient of the correlation length 
varies by 3\%, 4\%, 3\% respectively. Other quantities vary similarly.
\end{abstract}

\pacs{11.10.Gh; 11.10.Kk; 11.15.Ha; 12.38.Bx; 75.10.Hk}

\vskip 3mm

The two-dimensional nonlinear $\sigma$--model has been extensively 
studied since it is supposed to share with four-dimensional QCD
the property of being asymptotically free. A lot of work has been 
devoted to the checking of renormalization-group predictions. 
In order to make accurate comparisons it is very important to have 
high-order perturbative predictions. In \cite{carpel}
the renormalization-group $\beta$- and $\gamma$-functions 
were computed for the standard action with nearest-neighbour 
interactions on a square lattice. The result was expressed in terms 
of a small number of basic lattice integrals that were computed 
numerically. In Ref.~\cite{dds}, using the coordinate space method
\cite{Luescher-Weisz}, 
the lattice integrals that appear
in the results of \cite{carpel} were independently evaluated 
with higher numerical precision, and two of them were found 
grossly incorrect. As we shall discuss below, the large discrepancy 
found for one of them was due to the fact that its definition 
was not correctly given in Ref. \cite{carpel}.

Two of the present authors (B. A. and M. P.) 
have performed a complete independent check of the 
analytic part of the lattice calculation, by computing the 
two-point function in the presence of an external constant 
magnetic field to three loops. Their result is in agreement 
with that presented in Ref. \cite{carpel}. However their computation
allowed the discovery of an incorrect definition in Ref. \cite{carpel}:
the constant $W_2$ 
which was used in the computation of Ref. \cite{carpel} does not 
have the definition given in App. A.1.
Let us indicate with  $\widehat{W}_2$
the definition appearing in App. A.1 of Ref. \cite{carpel}:
\begin{eqnarray}
\widehat{W}_2 &\equiv & \lim_{h\to0} \left[
\int_{-\pi}^{+\pi}  {{{\rm d^2} q} \over  \left(2 \pi\right)^2}
\int_{-\pi}^{+\pi}  {{{\rm d^2} r} \over  \left(2 \pi\right)^2}
\int_{-\pi}^{+\pi}  {{{\rm d^2} s} \over  \left(2 \pi\right)^2}
\int_{-\pi}^{+\pi}  {{{\rm d^2} t} \over  \left(2 \pi\right)^2}
\left(2 \pi\right)^2 \delta^2(q+r+s+t) \right. \nonumber \\
 && \qquad\qquad\qquad \times {{\sum_{\mu\nu} \sin q_\mu\; \sin r_\nu\; 
     \sin s_\mu\; \sin t_\nu } \over
  \left( \widehat{q}^2 + h\right) \; \left( \widehat{r}^2 + h\right) \;
  \left( \widehat{s}^2 + h\right) \; \left( \widehat{t}^2 + h\right) \;
  \left( \widehat{q+r}^2 + h\right) \;} \; 
\nonumber \\
 && \left. - 
       {1\over 6} I(h)^3 + {1\over 8}\left(1 - {1\over \pi}\right)
       I(h)^2 - \left( {1\over 32} + {1\over 16\pi^2} -
       {1\over 16\pi} - {1\over 2} R\right) I(h) \right] \;,
\label{laequacion}
\end{eqnarray}
where $\widehat{p}^2 =\sum_\mu 4\sin^2 p_\mu/2$, $I(h)$ is 
the integral of the propagator
\begin{equation}
 I(h)\equiv 
\int_{-\pi}^{+\pi}  {{{\rm d^2} q} \over  \left(2 \pi\right)^2}
 {1\over \widehat{p}^2 + h } = -{1\over 4\pi} 
   \log \left({h \over 32}\right) + O(h)\; ,
\end{equation}
and $R\approx 0.0148430$ is a numerical constant 
introduced in~\cite{carpelviejo}. In the calculation of Ref. \cite{carpel}, 
$W_2$ was instead defined as 
\begin{equation}
W_2 = \widehat{W}_2 + {85\over 2304\pi^3} \zeta(3).
\end{equation}
This difference explains most of the discrepancy found in Ref. \cite{dds}
for the numerical estimate of $W_2$. Using \cite{dds}
\begin{equation}
\widehat{W}_2 = 0.0006923019(1),
\end{equation}
we find 
\begin{equation}
W_2 =\, 0.002122552,
\label{W2-new-estimate}
\end{equation}
which should be compared with $W_2 = 0.00221$ reported in 
Ref. \cite{carpel}. The remaining small discrepancy was due to 
numerical problems that were not understood at the time. 
We have completely revised the programs and now we find $W_2 = 0.002122(1)$
in agreement with (\ref{W2-new-estimate}). 
It follows that, while the 
analytic expressions of Ref. \cite{carpel} are correct, the numerical 
results for the perturbative corrections presented in 
Refs. \cite{carpel,dds} must be revised. Below we give the 
numerical expressions for the perturbative constants $a_2$, $b_2$, $c_3$,
and $b_2^{(n)}$ (see Ref. \cite{carpel} for definitions),
using the numerical estimates of Ref. \cite{dds} 
and the estimate (\ref{W2-new-estimate}):
\begin{eqnarray}
a_2 &=& {1\over (N-2)^2} (0.0444 + 0.0216N + 0.0045N^2 - 0.0129N^3),\\
b_2 &=& {1\over (N-2)^2} (0.1316 + 0.0187N - 0.0202N^2 - 0.0108N^3),\\
c_3 &=& {N-1\over (N-2)^3} (0.0121 + 0.0122N - 0.0070N^2 - 0.0092N^3 + 
          0.0041N^4),\\
b_2^{(n)} &=& {1\over (N-2)^2} 
   \left[ 0.0912 + 0.0446N + 0.0093N^2 - 0.0257N^3 \right.
\nonumber \\
     && \left. \qquad + n(n+N-2) (-0.0363 - 0.0187N+ 0.0149N^2) 
                      + 0.0041\, n^2 (n+N-2)^2\right].
\end{eqnarray}
For $N=3$ we have $a_2=-0.1969$, $b_2=-0.2853$, $c_3=0.1346$:
with respect to the previous estimates of Ref. \cite{carpel},
$a_2$, $b_2$, and $c_3$ vary by 3\%, 5\%, and 1\% respectively. 
The difference is very small and does not change the conclusions 
of the papers that compared Monte Carlo results with the 
perturbative predictions \cite{o3_letter,CEMPS,Alles-esoci,TheBest,Monaco}.

\end{document}